\documentclass[12pt ]{article}

\renewcommand{\theequation}
{\arabic{section}.\arabic{equation}}

\makeatletter
\def\eqnarray{ \stepcounter{equation} \let\@currentlabel=\theequation
 \global\@eqnswtrue
 \global\@eqcnt\z@
 \tabskip\@centering
 \let\\=\@eqncr
 $$\halign to \displaywidth\bgroup\@eqnsel\hskip\@centering
 $\displaystyle\tabskip\z@{##}$&\global\@eqcnt\@ne
 \hfil$\displaystyle{{}##{}}$\hfil
 &\global\@eqcnt\tw@$\displaystyle\tabskip\z@{##}$\hfil
 \tabskip\@centering&\llap{##}\tabskip\z@\cr}
\makeatother

\makeatletter
\def\@arrayacol{\edef\@preamble{\@preamble \hskip .5\arraycolsep}}
\def\array{\let\@acol\@arrayacol \let\@classz\@arrayclassz
\let\@classiv\@arrayclassiv \let\\\@arraycr\def\@halignto{}\@tabarray}
\makeatother

\makeatletter
\newcounter{subeqncnt}
\def\thesubeqncnt{\alph{subeqncnt}}
\def\subequations{\begingroup%
   \stepcounter{equation}\edef\@tempa{\theequation}%
   \let\c@equation\c@subeqncnt\c@subeqncnt\z@
   \edef\theequation{\@tempa\noexpand\thesubeqncnt}}

\makeatother

\newcommand{\be}{\begin{equation}}
\newcommand{\ee}{\end{equation}}

\newcommand{\beqa}{\begin{eqnarray}}
\newcommand{\eeqa}{\end{eqnarray}}
\newcommand{\nn}{\nonumber}

\newcommand{\eqref}[1]{(\ref{#1})}

\begin{document}

\setlength{\baselineskip}{7mm}
\begin{titlepage}
\begin{flushright}

{\tt NRCPS-HE-58-2013} \\

\end{flushright}

\begin{center}
{\Large ~\\{\it  Proton Structure \\
and  \\
Tensor Gluons
\vspace{1cm}

}

}

\vspace{1cm}

{\sl George Savvidy

\bigskip
\centerline{${}$ \sl Institute of Nuclear and Particle Physics}
\centerline{${}$ \sl Demokritos National Research Center, Ag. Paraskevi,  Athens, Greece}
\bigskip

}
\end{center}
\vspace{30pt}

\centerline{{\bf Abstract}}
We consider a possibility that inside the proton and, more generally, inside the hadrons there are
additional partons - tensor-gluons, which can carry a part of the proton momentum.  The tensor-gluons have
zero electric charge, like gluons, but have a larger spin. Inside the proton
a nonzero density of the tensor-gluons can be generated by the emission of tensor-gluons
by gluons. The last mechanism is typical for non-Abelian tensor gauge
theories, in which there exists a gluon-tensor-tensor vertex of order g.
Therefore the number of gluons changes not only
because a quark may radiate a gluon or because a gluon may split into a quark-antiquark
pair or into two gluons, but also because a gluon can split into two tensor-gluons.
The process of gluon splitting suggests that part of the proton momentum
which was carried by neutral partons is shared between vector and tensor gluons.
We derive evolution equations for the parton distribution functions  which take into account these new processes. The
momentum sum rule allows to find the tensor-gluons contribution to the Callan-Simanzik beta function
and to calculate the corresponding anomalous dimensions. This contribution
changes the behavior of the structure functions, and
the logarithmic correction to the Bjorken scaling becomes more mild. This also influences the unification scale
at which the coupling constants of the Standard Model merge,
shifting its value to lower energies of order of 40 TeV.

\vspace{12pt}

\noindent

\end{titlepage}



\pagestyle{plain}

\section{\it Introduction}

It was predicted that the Bjorken scaling should be broken by logarithms
of a transverse momentum $Q^2$ and that these
deviations from the scaling law can be computed for the deep inelastic
structure functions \cite{Gross:1973id,Politzer:1973fx,Altarelli:1977zs,Dokshitzer:1977sg,Gribov:1972ri,Gribov:1972rt,
Lipatov:1974qm,Fadin:1975cb,Kuraev:1977fs,Balitsky:1978ic,Cabibbo:1978ez,Gribov:1981ac}.
In the leading logarithmic approximation the results
can be formulated in the parton language \cite{Bjorken:1969ja} by assigning the well determined $Q^2$
dependence to the parton densities \cite{Gross:1973id,Politzer:1973fx,Gross:1973ju,Gross:1974cs,Altarelli:1977zs}.

In this article we shall  consider a possibility that inside the proton and, more generally,
inside the hadrons there are
additional partons - tensor-gluons, which can carry a part of the proton momentum.  Tensor-gluons have
zero electric charge, like gluons, but have a larger spin. Inside the proton
a nonzero density of the tensor-gluons can be generated by the emission of tensor-gluons
by gluons \cite{Savvidy:2005fi,Savvidy:2005zm,Savvidy:2005ki,Savvidy:2010vb}.
The last mechanism is typical for non-Abelian tensor gauge
theories, in which there exists a gluon-tensor-tensor vertex of order g
\cite{Savvidy:2005fi,Savvidy:2005zm,Savvidy:2005ki,Savvidy:2010vb}.
Therefore the number of gluons changes not only
because a quark may radiate a gluon or because a gluon may split into a quark-antiquark
pair or into two gluons \cite{Gross:1973ju,Gross:1974cs,Altarelli:1977zs},
but also because a gluon can split into two tensor-gluons
\cite{Savvidy:2005fi,Savvidy:2005zm,Savvidy:2005ki,Savvidy:2010vb,Georgiou:2010mf,Antoniadis:2011rr}.
The process of gluon splitting into tensor-gluons suggests that part of the proton momentum
which was carried by neutral partons is shared between vector and tensor gluons.

The proposed model was formulated in terms of field theory Lagrangian
\cite{Savvidy:2005fi,Savvidy:2005zm,Savvidy:2005ki,Savvidy:2010vb}.
The Lagrangian describes the interaction of the gluons with their massless partners of higher
spin \cite{Savvidy:2005fi,Savvidy:2005zm,Savvidy:2005ki,Savvidy:2010vb}. We shall call them tensor-gluons.
The characteristic property of the model is that all interaction vertices
between gluons and tensor-gluons have {\it dimensionless coupling constants} in four-dimensional space-time.
That is, the cubic interaction vertices have only first order derivatives and the quartic vertices
have no derivatives at all. These are familiar properties of the standard Yang-Mills field theory.
In order to understand the physical properties of the
model it was important to study the tree level scattering amplitudes.
A very powerful spinor helicity technique \cite{Berends:1981rb,Kleiss:1985yh,Xu:1986xb,
Gunion:1985vca,Dixon:1996wi,Parke:1986gb,Berends:1987me,
Witten:2003nn,Cachazo:2004by,Cachazo:2004dr,Britto:2004ap,Britto:2005fq,
Benincasa:2007xk,Cachazo:2004kj,Georgiou:2004by,
ArkaniHamed:2008yf,Berends:1988zn,Mangano:1987kp}
was used to calculate  high order tree level diagrams with the participation of tensor-gluons
in \cite{Georgiou:2010mf}.

These tree level scattering amplitudes describe a fusion of gluons into tensor-gluons
\cite{Georgiou:2010mf}.
They are generalizations of the Parke-Taylor scattering amplitude to the case
of two tensor gauge bosons of spin $s$ and $(n-2)$ gluons.
The result reads \cite{Georgiou:2010mf}:
\be\label{GSamplitude}
M_n(1^+,..i^-,...k^{+s},..j^{-s},..n^+)= g^{n-2}
 \frac{<ij>^4}{\prod_{r=1}^{n} <r (r+1)>} \Big( \frac{<ij>}{<ik>}\Big)^{2s-2},
\ee
where $n$ is the total number of particles and the dots stand for the
positive-helicity gluons. Here $i$ is the position of the negative-helicity gluon,
while $k$ and $j$ are the positions of the particles with helicities $+s$ and $-s$ respectively.
For $s=1$ this expression reduces  to the well-known result for
the MHV amplitude \cite{Parke:1986gb}.
The scattering amplitudes (\ref{GSamplitude}) were used to derive the amplitudes of splitting
of gluon into tensor-gluons in \cite{Antoniadis:2011rr}.

Here we shall use the splitting amplitudes in order to derive the generalization of the DGLAP evolution equations
for the parton distribution functions, which take into account the processes of
emission of tensor-gluons by gluons. The
momentum sum rule allows to find the tensor-gluons contribution to the one-loop
Callan-Simanzik beta function, which takes the following form:
\be\label{fullbetaint}
\alpha(t)= {\alpha \over 1+ b  \alpha ~t }~,~~~~~b=  {\sum_s(12s^2 -1) C_2(G) - 4 n_f T(R) \over 12 \pi},~~~~
\ee
where $s$ is the spin of the gauge bosons. The matrix of
anomalous dimensions for the twist-two operators $\gamma_n$ can also be computed. The spin-dependent
term in the Callan-Simanzik beta function coefficient (\ref{fullbetaint})
changes the $t= \ln(Q^2/Q^2_0)$ behavior of the structure functions
demonstrating that if the tensor-gluons exist,
the logarithmic correction to the Bjorken scaling changes.
This also influences the unification scale at which the coupling constants of the
Standard Model merge  \cite{Georgi:1974sy,Georgi:1974yf},
shifting its value to lower energies of order of $M \sim 40~ TeV$:
\be
\ln{M\over \mu} = {\pi \over 58} \left({1\over \alpha_{el}(\mu)}- {8\over 3} {1\over \alpha_s(\mu)}\right),
\ee
where $\alpha_{el}(\mu)$ and $\alpha_s(\mu)$ are the electromagnetic and strong coupling constants at
scale $\mu$.

The present paper is organized as follows. In section two the basic formulae for
scattering amplitude and splitting functions are recalled, definitions  and notations
are specified. In section three the generalized evolution equations that describe the
$Q^2$ dependence of parton densities is derived and the physical interpretation of the new terms
is presented. In section four we obtain the one-loop
Callan-Simanzik beta function for tensor-gluons and the anomalous dimensions for the
singlet and nonsiglet scaling functions. In section five we discuss the
unification of coupling constants of the Standard Model including one-loop contribution of the tensor-bosons.
Section six contains concluding remarks and the Appendix - the details of the regularization scheme.

\section{\it Splitting Functions}

It is convenient to  represent a scattering amplitude for the
massless particles of momenta $p_i$ and polarization tensors $\varepsilon_i$ ~$(i=1,...,n)$,
which are described by irreducible massless representations of the Poincar\'e group and are
classified by their helicities $h= \pm s$,  in the following form:
$$
M_n = M_n(p_1,\varepsilon_1;~p_2,\varepsilon_2;~...;~p_n,\varepsilon_n).
$$
Further by representing the momenta $p_i$ and polarization tensors $\varepsilon_i$ in terms of spinors
the scattering amplitude   $M_n$ can  be considered
as a function of spinors $\lambda_i$, $\tilde{\lambda}_i$ and helicities $h_i$
\cite{Berends:1981rb,Kleiss:1985yh,Xu:1986xb,
Gunion:1985vca,Dixon:1996wi,Parke:1986gb,Berends:1987me,
Witten:2003nn,Cachazo:2004by,Cachazo:2004dr,Britto:2004ap,Britto:2005fq,
Benincasa:2007xk,Cachazo:2004kj,Georgiou:2004by,
ArkaniHamed:2008yf,Berends:1988zn,Mangano:1987kp}:
\be\label{smatrix}
M_n=M_n(\lambda_1,\tilde{\lambda}_1,h_1;~...;~\lambda_n,\tilde{\lambda}_n,h_n)~.
\ee
The advantage of the spinor representation is that introducing a complex deformation of
the particles momenta
one can derive a general form of the three-particle interaction vertices
\cite{Benincasa:2007xk,Georgiou:2010mf}:
$$
M_3(1^{h_1} ,2^{h_2},3^{h_3} ).
$$
The dimensionality
of the three-point  vertex  $M_3(1^{h_1} ,2^{h_2},3^{h_3} )$  is $[mass]^{D=\pm(h_1+h_2+h_3)}$.

In the generalized Yang-Mills theory \cite{Savvidy:2005fi,Savvidy:2005zm,Savvidy:2005ki,Savvidy:2010vb}
all interaction vertices
between high-spin particles have {\it dimensionless coupling constants},
which means that the helicities of the interacting particles in the vertex are
constrained  by the relation $D=\pm(h_1+h_2+h_3)= 1$.
Therefore the interaction vertex between
massless tensor-gluons, the TTT-vertex,   has the following form \cite{Georgiou:2010mf}:
\beqa\label{dimensionone1}
M_3 &=& g f^{abc} <1,2>^{-2h_1 -2h_2 -1} <2,3>^{2h_1 +1} <3,1>^{2h_2 +1},~~~~h_3= -1 - h_1 -h_2, \nn\\
M_3 &=& g f^{abc} [1,2]^{2h_1 +2h_2 -1} [2,3]^{-2h_1 +1} [3,1]^{-2h_2 +1},~~~~~~~~~~~~~~~~~~h_3= 1 - h_1 -h_2,
\eeqa
where $f^{abc}$ are the structure constants of the internal gauge group G.
In particular, considering the interaction between a gluon of helicity $h_1 = \pm 1$ and
a tensor-gluon of helicity $h_2 = \pm s$, the GTT-vertex,  one can find from  (\ref{dimensionone1})  that
\be\label{vertecies}
h_3 = \pm \vert s-2 \vert,~ \pm s,~
 \pm \vert s+2 \vert~
\ee
and the corresponding gluon-tensor-tensor  interaction vertices GTT have
the form
\beqa\label{1ssvertex}
M^{a_1a_2a_3}_3(1^{-s} ,2^{-1} ,3^{+s} )&=& g ~f^{a_1 a_2 a_3} {<1,2>^{4} \over <1,2> <2,3> <3,1>}
\left({<1,2>  \over  <2,3> }\right)^{2s-2},\nn\\
M^{a_1a_2a_3}_3(1^{-s} ,2^{+1},3^{s-2} )&=& g ~f^{a_1 a_2 a_3} {<1,3>^{4} \over <1,2> <2,3> <3,1>}
\left({<1,2>  \over  <2,3> }\right)^{2s-2}.
\eeqa
These are the vertices which reduce to the standard triple gluon vertex  when $s=1$.
Using these vertices  one can compute the scattering amplitudes of gluons and  tensor-gluons.
The color-ordered
scattering amplitudes involving two tensor-gluons of helicities $h =\pm s$,
one negative helicity gluon
and $(n-3)$ gluons of positive helicity were found  in  \cite{Georgiou:2010mf}:
\be\label{gs}
\hat{M}_n(1^+,..i^-,...k^{+s},..j^{-s},..n^+)=i g^{n-2} (2\pi)^4 \delta^{(4)}(P^{a\dot{b}})
 \frac{<ij>^4}{\prod_{l=1}^{n} <l l+1>} \Big( \frac{<ij>}{<ik>}\Big)^{2s-2},
\ee
where $n$ is the total number of particles and the dots stand for any number of
positive helicity gluons, $i$ is the position of the negative-helicity gluon,
while $k$ and $j$ are the positions of the gluons with helicities $+s$ and $-s$ respectively.
The expression \eqref{gs}  reduces to the
famous Parke-Taylor formula \cite{Parke:1986gb} when $s=1$.
In particular, the five-particle amplitude takes the following form:
\beqa\label{fivepointamplitude}
\hat{M}_5(1^+,2^-,3^{+},4^{+s},5^{-s})=  i g^{3} (2\pi)^4 \delta^{(4)}(P^{a\dot{b}})
\frac{<25>^4}{\prod_{i=1}^{5}
<i i+1>}  (\frac{<25> }{<24>})^{2s-2},
\eeqa
where
$
P^{a\dot{b}} = \sum^n_{m=1} \lambda^a_m \tilde{\lambda}^{\dot{b}}_m
$ is the total momentum. The scattering amplitudes (\ref{gs}) and (\ref{fivepointamplitude}) can be used to extract splitting amplitudes
of gluons and tensor-gluons \cite{Antoniadis:2011rr}. The collinear behavior
of the tree amplitudes has the following factorized  form
\cite{Dixon:1996wi,Parke:1986gb,Berends:1987me,Berends:1988zn,Mangano:1987kp}:
\be\label{factorization}
M^{tree}_n(...,a^{\lambda_a},b^{\lambda_b},...)~~  {a \parallel b \over \rightarrow}  ~~  \sum_{\lambda=\pm 1}
Split^{ tree }_{-\lambda}(a^{\lambda_a},b^{\lambda_b})~ \times ~M^{tree}_{n-1}(...,P^{\lambda} ,...),
\ee
where $Split^{tree}_{-\lambda}(a^{\lambda_a},b^{\lambda_b})$ denotes the splitting amplitude and the
intermediate state $P$ has momentum $k_P=k_a +k_b$ and helicity $\lambda$.
Considering the amplitude \eqref{fivepointamplitude} in the limit when the
particles 4 and 5 become collinear, $k_4  \parallel k_5$, that is,
$k_4 = z k_P,~k_5 = (1-z) k_P$,  $k^2_P \rightarrow 0$ and $z$ describes the
longitudinal momentum sharing, one can deduce that the corresponding behavior of spinors is
$
\lambda_4 = \sqrt{z} \lambda_P,~~~\lambda_5 = \sqrt{1-z} \lambda_P,
$
and  that the amplitude \eqref{fivepointamplitude} takes the following factorization
form \cite{Antoniadis:2011rr}:
\beqa
M_5(1^+,2^-,3^{+},4^{+s},5^{-s})
 = A_4(1^+,2^-,3^{+},P^-) \times ~ Split_+(a^{+s},b^{-s}),\nn
\eeqa
where
\be\label{spliting1}
Split_+(a^{+s},b^{-s}) = \left(\frac{1-z}{z } \right)^{ s-1}  \frac{(1-z)^2}{\sqrt{z(1-z)}}
\frac{1}{ <a, b>}.
\ee
In a similar way one can deduce that
\be\label{spliting2}
Split_+(a^{-s},b^{+s}) = \left(\frac{z }{1-z} \right)^{ s-1}  \frac{z^2}{\sqrt{z(1-z)}}
\frac{1}{ <a, b>}.
\ee
Considering different collinear limits $k_1  \parallel k_5$ and $k_3  \parallel k_4$
one can get \cite{Antoniadis:2011rr}
\be\label{spliting3}
Split_{+s}(a^{+},b^{-s}) =  \frac{(1-z)^{s+1}}{\sqrt{z(1-z)}}
\frac{1}{ <a, b>},~~~
Split_{+s}(a^{-s},b^{+}) =  \frac{z^{s+1}}{\sqrt{z(1-z)}}
\frac{1}{ <a, b>}
\ee
and
\be\label{spliting4}
Split_{-s}(a^{+s},b^{+}) =  \frac{z^{-s+1}}{\sqrt{z(1-z)}}
\frac{1}{ <a, b>}, ~~~
Split_{-s}(a^{+},b^{+s}) =    \frac{(1-z)^{-s+1}}{\sqrt{z(1-z)}}
\frac{1}{ <a, b>}.
\ee
The set of splitting amplitudes (\ref{spliting1})-(\ref{spliting4})
 $G\rightarrow TT$, $T \rightarrow GT$ and
$T \rightarrow TG$ reduces to the full set of gluon splitting amplitudes
\cite{Dixon:1996wi,Parke:1986gb,Berends:1987me,Berends:1988zn,Mangano:1987kp}
when $s=1$.

Since the collinear limits of the scattering amplitudes
are responsible for parton evolution  \cite{Altarelli:1977zs}
we can extract from the above expressions the
Altarelli-Parisi splitting probabilities for
tensor-gluons . Indeed, the residue of the collinear
pole in the square (of the factorized
amplitude (\ref{factorization})) gives Altarelli-Parisi splitting probability $P(z)$:
\be\label{AltarelliParisi}
P(z)= C_2(G) \sum_{h_P , h_a, h_b} \vert Split_{-h_P}(a^{h_a},b^{h_b}) \vert^2 ~ s_{ab}.
\ee
where $s_{ab}=2 k_a \cdot k_b= <a,b>[a,b]$.
The invariant operator $C_2$ for the representation R is defined by the equation
$ t^a t^a  = C_2(R)~ 1 $ and $tr(t^a t^b) = T(R) \delta^{ab}$.
Substituting the splitting amplitudes (\ref{spliting1})-(\ref{spliting4})
into (\ref{AltarelliParisi}) we are getting
\beqa\label{setoftensorgluon}
P_{TG}(z) &=&  C_2(G)\left[ {z^4 \over z(1-z)}\left( {z\over 1-z} \right)^{2s-2}
+{(1-z)^4 \over z(1-z)} \left( {1-z\over  z}\right)^{2s-2} \right],\nn\\
P_{GT}(z) &=&  C_2(G)\left[ {1\over z(1-z)}\left( {1\over 1-z} \right)^{2s-2}
+{(1-z)^4 \over z(1-z)} (1-z)^{2s-2} \right],\\
P_{TT}(z) &=&  C_2(G)\left[ {z^4 \over z(1-z)} z^{2s-2}
+{1 \over z(1-z)} \left( {1\over  z}\right)^{2s-2} \right]. \nn
\eeqa
The momentum conservation in the vertices clearly fulfils because these
functions satisfy the relations
\be
P_{TG}(z)=P_{TG}(1-z),~~~P_{GT}(z)=P_{TT}(1-z),~~~~~~~z < 1.
\ee
In the leading order the kernel $P_{TG}(z)$ has a meaning of variation per unit t of the probability
density of finding a tensor-gluon inside the gluon, $P_{GT}(z)$ - of finding gluon inside
the tensor-gluon and $P_{TT}(z)$ - of finding tensor-gluon inside the tensor-gluon.
For completeness we shall present also quark and gluon splitting functions \cite{Altarelli:1977zs}:
\beqa\label{setofquarkgluon}
P_{qq}(z) &=& C_2(R){1+z^2 \over 1-z },\nn\\
P_{Gq}(z) &=& C_2(R){1+(1-z)^2 \over z },\\
P_{qG}(z) &=& T(R)[z^2 +(1-z)^2], \nn\\
P_{GG}(z) &=&  C_2(G)\left[{1 \over z(1-z)}+ {z^4 \over z(1-z)}+{(1-z)^4 \over z(1-z)}\right],\nn
\eeqa
where $C_2(G)= N, C_2(R)={N^2-1  \over  2 N},  T(R) = {1  \over  2}$ for the SU(N) groups.

We have to notice that the limit $s \rightarrow 1/2$ in
the scattering amplitudes (\ref{gs}) and (\ref{fivepointamplitude})
reduces them to the tree level gluon scattering amplitudes into a
quark pair.
Thus the formula has larger validity area than the area in which it has been initially
derived \cite{Georgiou:2010mf}. This fact should be visible in the splitting probabilities
as well. Indeed let us consider the tensor splitting functions (\ref{setoftensorgluon}) and
take the limit to the half-integer spin $s \rightarrow 1/2$. One can see  that
tensor-gluon splitting probabilities  (\ref{setoftensorgluon}) reduce to the
quark-gluon splitting probabilities  (\ref{setofquarkgluon})
\be
P_{TT}(z)  \rightarrow P_{qq}(z),~~~P_{GT}(z)  \rightarrow P_{Gq}(z),~~~
P_{TG}(z)  \rightarrow P_{qG}(z).
\ee
Taking into account the remark which we made after the formula (\ref{spliting4})
one can see that
the substitution $s \rightarrow 1$ gives gluon-gluon splitting function as well
\be
{1\over 2}(P_{TG}(z) +P_{GT}(z)  + P_{TT}(z) ) \rightarrow  P_{GG}(z).
\ee

Having in hand the new set of splitting probabilities
for tensor-gluons (\ref{setoftensorgluon}) we can hypothesize that
a possible emission of tensor-gluons
should produce a tensor-gluon "shower"  of neutral partons
inside the proton  in additional to the quark and gluon "shower'.
Our goal is to derive
DGLAP equations \cite{Altarelli:1977zs,Gribov:1972ri,Gribov:1972rt,Lipatov:1974qm,Fadin:1975cb,Kuraev:1977fs,
Balitsky:1978ic,Dokshitzer:1977sg} which will take into account these new emission processes.

\section{\it Generalization of DGLAP Equation}

It is well known that the deep inelastic structure functions
can be  expressed in terms of parton quark densities.
If $q^i(x)$ is the density of quarks of type i (summed over
colors) inside a proton target with fraction x of the proton longitudinal momentum
in the infinite momentum frame then the scaling structure functions can
be represented in the form
$$
2F_1(x)= F_2(x)/x= \sum_i Q^2_i [q^i(x)+\bar{q}^i(x)].
$$
The scaling behavior of the structure functions is broken and the results
can be formulated by assigning  a well determined $Q^2$
dependence to the parton densities.  This can be achieved by
introducing the integro-differential equations which describe the
$Q^2$ dependence of quark  $q^i(x,t)$ and gluon densities  $G(x,t)$,
where $t=\ln(Q^2/Q^2_0)$ \cite{Altarelli:1977zs,Gribov:1972ri,Gribov:1972rt,
Lipatov:1974qm,Fadin:1975cb,Kuraev:1977fs,
Balitsky:1978ic,Dokshitzer:1977sg}.

Let us see what will happen if one supposes that
there are additional partons - tensors-gluons - inside the proton.
In accordance with our hypothesis there
is  an additional emission of tensor-gluons in the proton, therefore one should introduce
the corresponding density $T(x, t)$ of tensor-gluons (summed over colors)
inside the proton in the $P_{\infty}$ frame. We can  derive
the integro-differential equations that describe the $Q^2$ dependence
of parton densities in this general case. They are:
\beqa\label{evolutionequation}
{d q^i(x,t)\over dt} &=& {\alpha(t) \over 2 \pi} \int^{1}_{x} {dy \over y}[\sum^{2 n_f}_{j=1} q^j(y,t)~
P_{q^i q^j}({x \over y})+ G(y,t)~ P_{q^i G}({x \over y})] ,\\
{d G(x,t)\over dt} &=& {\alpha(t) \over 2 \pi} \int^{1}_{x} {dy \over y}[\sum^{2 n_f}_{j=1} q^j(y,t)~
P_{G q^j}({x \over y})+ G(y,t) ~P_{G G}({x \over y})+ T(y,t) ~P_{G T}({x \over y}) ],\nn\\
{d T(x,t)\over dt} &=& {\alpha(t) \over 2 \pi} \int^{1}_{x} {dy \over y}[
G(y,t)~ P_{T G}({x \over y}) +  T(y,t)~ P_{T T}({x \over y})].\nn
\eeqa
The $\alpha(t)$ is the running coupling constant ($\alpha = g^2/4\pi$).
In the leading logarithmic approximation $\alpha(t)$ is of the form
\be\label{strongcouplingcons}
{\alpha \over \alpha(t)} = 1 +b ~\alpha ~t~~,
\ee
where $\alpha = \alpha(0)$ and $b$ is the one-loop Callan-Simanzik coefficient,
which, as we shall see below, receives an additional contribution from the tensor-gluons.
Here the indices i and j run over quarks and antiquarks of all flavors. The number
of quarks of a given fraction of momentum changes when a quark looses momentum by
radiating a  gluon,
or a  gluon inside the proton may produce a quark-antiquark pair \cite{Altarelli:1977zs}.
Similarly the number of  gluons changes
because a quark may radiate a gluon or because a gluon may split into a quark-antiquark
pair or into two gluons or {\it into two tensor-gluons}. This last possibility is realized,
because, as we have seen, in non-Abelian tensor gauge
theories there is a triple vertex GTT (\ref{1ssvertex}) of a gluon and two tensor-gluons of order g
\cite{Savvidy:2005fi,Savvidy:2005zm,Savvidy:2005ki,Savvidy:2010vb}.
This interaction should be taken into consideration,  and we added the
term $T(y,t) ~P_{G T}({x \over y})$ in the second
equation (\ref{evolutionequation}). The density of tensor-gluons $T(x,t)$  changes when
a gluon splits into two tensor-gluons or when a tensor-gluon radiates a gluon. This
evolution is described by the last equation (\ref{evolutionequation}).

In order to guarantee that the total momentum of the proton, that is, of
all partons is unchanged, one should impose the following constraint:
\be\label{conservation}
{d\over dt}\int_{0}^{1} dz z [\sum^{2n_f}_{i=1}q^{i}(z,t)+G(z,t)+T(z,t)]=0.
\ee
Using the evolution equations (\ref{evolutionequation}) one can express the derivatives
of the densities in (\ref{conservation}) in terms of kernels and to see that the following
momentum sum rules should be fulfilled:
\beqa\label{momentumsum}
&&\int_{0}^{1} dz z [P_{qq}(z)+P_{Gq}(z) ]=0,\nn\\
&&\int_{0}^{1} dz z [2 n_f P_{qG}(z)+P_{GG}(z)+P_{TG}(z)]=0,\nn\\
&&\int_{0}^{1} dz z [ P_{GT}(z)+P_{TT}(z)]=0.
\eeqa
Before analyzing these momentum sum rules let us first
inspect the behavior of the tensor-gluon kernels (\ref{setoftensorgluon})
at the end points $z=0,1$. As one can see,
they are singular at the boundary values similarly to the
case of the standard kernels (\ref{setofquarkgluon}).
Though there is a difference here: the singularities are of higher order compared to the standard case
\cite{Altarelli:1977zs}.
Therefore one should define the regularization procedure for the singular factors
$(1 - z)^{-2s+1}$ and $ z^{-2s+1}$  reinterpreting them as the  distributions $(1 - z)^{-2s+1}_{+}$ and
$z^{-2s+1}_{+}$, similar to the Altarelli-Parisi regularization \cite{Altarelli:1977zs}.
We shall define them  in the following
way:
\beqa\label{definition}
\int_{0}^{1} dz {f(z)\over (1 - z)^{2s-1}_+}&=&
\int_{0}^{1} dz {f(z)- \sum^{2s-2}_{k=0} {(-1)^k \over k!} f^{(k)}(1) (1-z)^k \over (1 - z)^{2s-1}},\nn\\
\nn\\
\int_{0}^{1} dz {f(z)\over z ^{2s-1}_+}&=&
\int_{0}^{1} dz {f(z)- \sum^{2s-2}_{k=0} {1 \over k!} f^{(k)}(0) z^k \over z^{2s-1}},\\
\nn\\
\int_{0}^{1} dz {f(z)\over z_+ (1-z)_+}&=&
\int_{0}^{1} dz {f(z)-  (1-z)f(0) - z f(1) \over z  (1-z) },\nn
\eeqa
where $f(z)$ is any test function which is sufficiently regular at the end points
and, as one can see, the defined substraction guarantees the convergence of the integrals.
Using the same arguments as in the standard case \cite{Altarelli:1977zs} we should add the delta function
terms into the definition of the diagonal kernels so that they will completely determine
the behavior of $P_{qq}(z)$ , $P_{GG}(z)$ and $P_{TT}(z)$ functions. The first equation
in the momentum sum rule (\ref{momentumsum})
remains unchanged because there is no tensor-gluon contribution into the quark evolution. The second
equation in the momentum sum rule (\ref{momentumsum}) will take the following form
(see Appendix for details):
\beqa\label{betacoefficient}
&\int_{0}^{1} dz z [2 n_f P_{qG}(z)+P_{GG}(z)+P_{TG}(z) + b_G \delta (z-1)]=\nn\\
&=\int_{0}^{1} dz z  [2 n_f T(R)[z^2 +(1-z)^2]+C_2(G)\left[{1 \over z(1-z)}
+ {z^4 \over z(1-z)}+{(1-z)^4 \over z(1-z)}\right]+\nn\\
&+C_2(G)\left[ {z^4 \over z(1-z)}\left( {z\over 1-z} \right)^{2s-2}
+{(1-z)^4 \over z(1-z)} \left( {1-z\over  z}\right)^{2s-2} \right]  ] +b_G=\nn\\
&={2\over 3} n_f T(R) - {11 \over 6}C_2(G)- {12 s^2 -1\over 6} C_2(G) + b_G =0.
\eeqa
From this result we can extract an additional contribution
to the one-loop Callan-Symanzik beta function
for gluons $b_{GG} $ arising from the tensor-gluon loop.
Indeed, the first beta-function coefficient enters into this expression because the momentum sum
rule (\ref{momentumsum}) implicitly comprises unitarity, thus the one-loop effects \cite{Altarelli:1977zs}.
In (\ref{betacoefficient}) we have three terms which come from quark loops:
\be
b^q_{GG} = -{  2 n_f   \over 3}T(R),
\ee
from the gluon loop:
\be\label{gluonsbeta}
b^G_{GG} =  {11 \over 6}C_2(G)
\ee
and  from the gauge boson loop of spin s:
\be\label{spinscontr}
b^T_{GG} =  {12 s^2 -1\over 6} C_2(G), ~~~s=1,2,3,4,....
\ee
It is a very interesting result because at s=1 we are rediscovering the
asymptotic freedom result \cite{Gross:1973id,Politzer:1973fx}.
For larger spins the  tensor-gluon contribution into the Callan-Simanzik beta function
has the same signature as the standard gluons, which means that tensor-gluons
"accelerate" the asymptotic freedom (\ref{strongcouplingcons}) of the strong
interaction coupling constant $\alpha(t)$.
The contribution is increasing quadratically
with spin of the tensor-gluons, that is, at large transfer momentum
the strong coupling constant tends to zero faster compared to the standard case:
\be
\alpha(t)= {\alpha \over 1+ b  \alpha ~t }~,
\ee
where
\be\label{fullbeta}
b =  {\sum_s(12s^2 -1) C_2(G) - 4 n_f T(R) \over 12 \pi},~~~~~s=1,2,...
\ee
In particular, the presence of the spin-two tensor-gluons in the proton will give
\be\label{fullbeta2}
b=  {58 C_2(G) - 4 n_f T(R) \over 12 \pi}.
\ee
Surprisingly, a similar result based on the parametrization of the
charge renormalization taken in the form  $b = (-1)^{2s}(A+Bs^2)$
was conjectured  by Curtright
\cite{Curtright:1981wv}. Here $A$ represents an orbital contribution and
$B s^2$ - the anomalous magnetic moment contribution
\cite{Savvidy:1977as,Matinyan:1976mp,Batalin:1976uv}. The unknown coefficients A and B were found
by comparing the suggested parametrization with the known results for s= 0, 1/2 and 1.

It is also possible to consider a straitforward generalization of the result obtained
for the effective action in Yang-Mills theory long ago
\cite{Savvidy:1977as,Matinyan:1976mp,Batalin:1976uv,Kay:1983mh}
to the higher spin gauge bosons. With the spectrum of the tensor-gluons in
the external chromomagnetic field  $\lambda = (2n+1 + 2s)gH +k^2_{\parallel}$
one can perform a summation of the modes and get an exact result for the one-loop effective
action similar to \cite{Savvidy:1977as,Kay:1983mh}:
\be
\epsilon= {H^2 \over 2} +{(gH)^2 \over 4\pi} ~b ~[\ln{gH \over \mu^2}-{1\over 2}],
\ee
where
\beqa\label{savv}
b =  -{2 C_2(G)\over \pi} ~ \zeta(-1, {2s+1\over 2})=
{12s^2-1\over 12 \pi}C_2(G),
\eeqa
and $\zeta(-1, q)=-{1\over 2}(q^2 -q +{1\over 6})$ is the generalized zeta
function\footnote{The generalized zeta function is defined as $\zeta(p, q)=\sum^{\infty}_{k=0}{1\over (k+q)^p}
={1\over \Gamma(p)} \int^{\infty}_{0} dt t^{-1+p} { e^{-qt} \over 1-e^{-t}} $ .}.
Because the coefficient in front of the logarithm defines  the beta function
\cite{Savvidy:1977as,Matinyan:1976mp}, one can see that (\ref{savv}) is
in agrement with the result (\ref{spinscontr}).
It is also interesting to mention that the spectrum of the open strings in the background
magnetic field has a similar spectrum with the gyromagnetic ration equal to two,
as it was pointed out by Bachas and Fabre in \cite{Bachas:1996zt}.

The third equation in the momentum sum rule (\ref{momentumsum}) will take the following form:
\beqa\label{tensorbeta}
&\int_{0}^{1} dz z [P_{TT}(z) + P_{GT}(z) + b_T \delta (z-1)]=\nn\\
&= C_2(G)\int_{0}^{1} dz z    \left[ {z^{2s+1} \over  1-z }
+{1\over (1-z) ~z^{2s-1}} +{1\over  z ~(1-z)^{2s-1}} +  {(1-z)^{2s+1} \over   z } \right]
+ b_T  \Rightarrow \nn\\
&\Rightarrow ~~Reg = C_2(G)\int_{0}^{1} dz      \left[ {z^{2s+2}-1 \over  1-z }
  +  (1-z)^{2s+1}   \right]  + b_T =\nn\\
&= -C_2(G) \sum^{2s+1}_{j=1} {1 \over j}  + b_T =0.
\eeqa
And again we can extract  the one-loop coefficient of the Callan-Symanzik beta function now for
tensor-gluon of spin s, which has the form
\be
b_{TT}  = C_2(G) \sum^{2s+1}_{j=1} {1 \over j}, ~~~s=1,2,3,4,....
\ee
As one can see, at s=1 we have
$$
C_2(G)~(1+{1\over 2 } +{1\over 3}) = {11 \over 6}C_2(G),
$$
and it coincides with the one loop contribution of the gluons (\ref{gluonsbeta}). The coefficient
grows as $\ln s$. This growth  is slower than $s^2$ in (\ref{fullbeta}). The reason
is that, as we shall discuss at the end of this section, the splitting of tensor-gluons into
pair of tensor-gluons was discarded in the derivation of the evolution equation.
In this article we limit ourselves by considering only emissions which always involve
the standard gluons and lower-spin tensors, ignoring infinite "stair" of transitions
between tensor-gluons.

In summary, we have to add $\delta (z-1)$  to the diagonal  kernels $P_{qq}(z)$, $P_{GG}(z)$ and  $P_{TT}(z)$
with the coefficients which have been determined by using the momentum sum rule (\ref{momentumsum})
 guaranteeing   the total momentum conservation of a hadron:
\beqa
P_{qq}(z)&=& C_2(R)\left[ {1+z^2 \over (1-z)_+ } + {3\over 2}\delta (z-1)\right],\nn\\
P_{GG}(z) &=&  2 C_2(G)\left[{z \over (1-z)_+}+ {1-z  \over z }+ z(1-z) \right]
+  {\sum_s(12s^2 -1)C_2(G) - 4 n_f T(R)  \over 6}    ~ \delta (z-1),\nn\\
P_{TT}(z) &=&  C_2(G)\left[{z^{2s+1} \over (1- z)_+ }   + {1\over (1-z) z^{2s-1}_+}
+ \sum^{2s+1}_{j=1} {1 \over j} ~ \delta (z-1) \right].
\eeqa
Thus we completed the definition of the kernels appearing in the
evolution equations (\ref{evolutionequation}).

At the end of this section we shall discuss, what type of processes could also be included into
the evolution equations (\ref{evolutionequation}). In (\ref{evolutionequation})
we ignore contribution of
the high-spin fermions $\tilde{q}^i$ of spin $s + 1/2$, which are the partners of the
standard quarks \cite{Savvidy:2005fi,Savvidy:2005zm,Savvidy:2005ki,Savvidy:2010vb},
supposing  that they are even heavier than the top quark. That is,
all kernels $P_{q\tilde{q}},P_{G\tilde{q}}, P_{T\tilde{q}},P_{\tilde{q}\tilde{q}},P_{\tilde{q}q},
P_{\tilde{q}G},P_{\tilde{q}T}$ and $P_{Tq}$ with the emission of $\tilde{q}^i$ are taken to be
zero. These terms can be included in the case of very high energies, but at modern energies
it seems safe to ignore these contributions. In the evolution equation
for tensor-gluons in (\ref{evolutionequation}) one could also include the
kernels $P_{TT^{'}}$ which describe the emission of  tensor-gluons by tensor-gluons,
the $TT^{'}T^{''}$ vertex (\ref{dimensionone1}) \cite{Savvidy:2005fi,Savvidy:2005zm,Savvidy:2005ki,Savvidy:2010vb}.
That also can be done and the number of evolution equations in that case will
tend to infinity. In this article we shall limit ourselves considering only emissions
which always involve
the standard gluons and lower spin tensors, ignoring infinite "stair" of transitions
between tensor-gluons. We shall consider a more general case in a separate work.

It is also natural to ask what will happen if one takes into consideration the contribution
of tensor-gluons of all spins into the beta function\footnote{I would like to thank
John Iliopoulos and Constantin Bachas for rasing this question.}. One can suggest two scenarios.
In the first one the high spin gluons, let us say of $s  \geq 3$, will get large mass
and therefore can be ignored at a given energy scale. In the second case,
when all of them remain massless, then one can
suggest the Reimann zeta function regularization, similar to the Brink-Nielsen regularization
\cite{Brink:1973kn}, which gives:
\be
b = C_2(G)  \sum^{\infty}_{s=1} {( 12s^2 -1) \over 12 \pi}=
C_2(G)[{1\over \pi}  \zeta(-2)- {1\over 12\pi} \zeta(0)] ={1\over 24\pi}C_2(G),
\ee
where $\zeta(-2)=0,~ \zeta(0)=-1/2$ and the theory remains asymptotically free
with a smaller beta coefficient, but this summation requires further justification.

\section{\it Moments of the Scaling Structure Functions}
The moments of the scaling structure functions measure the Fourier transform of the
coefficients $C^{(n)}(Q^2,g)$ of the Wilson's operator-product expansion of currents
\cite{Gross:1974cs}:
$$
\int^{1}_{0} dx x^{n-1} F_{i}(x,Q^2) ~~~~~ \begin{array}{c}
       \\
  \sim \\
  {Q^2 \rightarrow \infty}
\end{array}
~~~~~ C^{(n)}_{i}(t) <n\vert O^{(n)}\vert n>.
$$
Their logarithmic deviations from Bjorken scaling are obtainable from calculation
of anomalous dimensions $\gamma_n$ of the corresponding  operators and beta function
coefficient $b$ in (\ref{fullbeta}):
$$
C^{(n)}_{i}(t) ~~~~~ =
~~~~~ C^{(n)}_{i}(0)~ \left[{\alpha \over \alpha(t)}\right]^{ {\gamma_n/ 2\pi b}}.
$$
Therefore one should calculate the matrix of relevant anomalous dimensions of the
twist-two operators. They are represented in the terms of moments of the kernels
in (\ref{evolutionequation}):
\beqa\label{anomamartix}
\gamma _n =\int^{1}_{0} dz z^{n-1}\left(\begin{array}{ccc}
  P_{qq}(z)&2 n_f P_{qG}(z)&0\\
  P_{Gq}(z)&P_{GG}(z)&P_{GT}(z)\\
 0&P_{TG}(z)&P_{TT}(z)
\end{array} \right)=
 \left(\begin{array}{ccc}
  \gamma^{qq}_n& 2 n_f \gamma^{qG}_n&0\\
  \gamma^{Gq}_n&\gamma^{GG}_n&\gamma^{GT}_n\\
 0&\gamma^{TG}_n&\gamma^{TT}_n
\end{array} \right),
\eeqa
where the corresponding integrals for quarks and gluons are well known \cite{Gross:1973id}:
\beqa\label{quarkgluonanom}
\int_{0}^{1} dz z^{n-1} P_{qq}(z)&=& C_2(R)\left[- {1 \over 2 } + {1\over n(n+1)}
- 2\sum^{n}_{j=2}{1\over j}\right],\nn\\
\int_{0}^{1} dz z^{n-1} P_{Gq}(z)&=& C_2(R)\left[  {2+n +n^2 \over n(n^2-1)}\right] , \\
 2 n_f \int_{0}^{1} dz z^{n-1} P_{qG}(z)&=&  2 n_f T(R) \left[  {2+n +n^2 \over n(n+1)(n+2)}\right],\nn\\
\int_{0}^{1} dz z^{n-1} P_{GG}(z) &=&   C_2(G) [- {1 \over 6 } + {2\over n(n-1)}
+ {2\over (n+1)(n+2)} -\nn\\
&&~~~~~~~ -  2\sum^{n}_{j=2}{1\over j}-
{ 2 n_f T(R)  \over 3  C_2(G)}   + {\sum_s (12s^2 -1) \over 6}].\nn
\eeqa
We have to calculate the new terms in the matrix of anomalous dimensions
which are the contribution of tensor-gluons.  For tensors-gluons kernel $P_{TG}(z)$ they are:
\beqa\label{quarkgluonanomt1}
\int_{0}^{1} dz z^{n-1} P_{TG}(z) &=&  C_2(G)\left[ \sum^{2s+n}_{k=2s-1}
 {  (-1)^k (2s+n)!  \over k!(2s+n -k)! }~
{1\over  k-2s+2 } \right] +\\
&&C_2(G)\left[\sum^{2s+1}_{k=2s-n}   {  (-1)^k (2s+1)!  \over k!(2s+1 -k)! }~
{1 \over   k +1+n- 2s}\right] ,~~~ n \leq 2s -1\nn\\
\int_{0}^{1} dz z^{n-1} P_{TG}(z) &=&  C_2(G)\left[ \sum^{2s+n}_{k=2s-1}   {  (-1)^k (2s+n)!  \over k!(2s+n -k)! }~
{1\over  k-2s+2 } \right] +\nn\\
&&C_2(G)\left[   {  \Gamma(2s+2)  \Gamma(n-2s+1)  \over \Gamma(n+3) } \right] ,
~~~~~~~~~~~~~~~~~~ n \geq 2s,\nn
\eeqa
for the gluons-tensors kernel $P_{GT}(z)$:
\beqa\label{quarkgluonanomt3}
\int_{0}^{1} dz z^{n-1} P_{GT}(z) &=&     C_2(G)\left[
{ \Gamma(2s+2) \Gamma(n-1)   \over \Gamma(n+2s+1)  }~\right]
  ,~~~~n \leq 2s \nn\\
\int_{0}^{1} dz z^{n-1} P_{GT}(z) &=&     C_2(G) [ \sum^{n-2}_{k=2s-1}
{  (-1)^k (n-2)!  \over k!(n -k-2)! }~
{1\over  k-2s+2 } +\\
&+&{ \Gamma(2s+2) \Gamma(n-1)   \over \Gamma(n+2s+1)  }~],~~~~~~~~~~~~~n \geq 2s+1\nn
\eeqa
and for the tensors-tensors kernel $P_{TT}(z)$:
\beqa\label{quarkgluonanomt2}
\int_{0}^{1} dz z^{n-1} P_{TT}(z) &=& - C_2(G) \sum^{n+2s}_{j=2+2s} {1\over j}    ,~~~~~~~~~~~~~~
n \leq 2s, \nn\\
\int_{0}^{1} dz z^{n-1} P_{TT}(z) &=& - C_2(G) \left[\sum^{n+2s}_{j=2+2s} {1\over j}+
\sum^{n-2s }_{j=1} {1\over j} \right] ,~~~~~~~~~~
n \geq 2s+1
\eeqa
where $s=2,3,4,...$. For the nonsinglet piece  of the structure functions the relevant
anomalous dimension is $\gamma^{qq}_n$, and it does not have any additional
contribution from tensor-gluons, therefore for nonsinglet pieces of the structure functions one can  get
\be
\int^{1}_{0} dx x^{n-2} F^{NS}(x,Q^2) ~~~~~
\begin{array}{c}
  \\
  \sim \\
  {Q^2 \rightarrow \infty}
\end{array}
~~~~~ C^{(n)}_{NS}~\left[{\alpha \over \alpha(t)}\right]^{A^{NS}_n},
\ee
where
\be
A^{NS}_n ={\gamma^{qq}_n\over 2 \pi  b} = -{3 C_2(R) \over \sum_s(12 s^2 -1)C_2(G) -4n_f T(R)}
\left[ 1 - {2\over n(n+1)}
+4\sum^{n}_{j=2}{1\over j}\right].
\ee
The difference between it and the standard case
comes from the value of the coefficient $b$ in the Callan-Simanzik
beta function (\ref{fullbeta}).
This result shows  that if tensor-gluons exist in the strong interaction,
the logarithmic correction to the Bjorken scaling becomes more mild.
For the $SU(3)_c$ ($ C_2(G)=3, C_2(R)=4/3, T(R)=1/2, n_f=3$) and spin-two gluons $s=2$  we are getting
\be
A^{NS}_2=- {2\over 63},~~A^{NS}_3= -{25\over 504},~~...~~,A^{NS}_n= -{1\over 21}\ln n~~
\ee
instead of $A^{NS}_2= -{16\over 81},~~A^{NS}_3= -{25\over 81},~~...~~,A^{NS}_n=-
{8\over 27}\ln n$ in the standard case \cite{Gross:1974cs}.

For the singlet pieces  of the structure functions one should take
into account the mixing of gluon, fermion and tensor operators, so that
\be
\int^{1}_{0} dx x^{n-2} F^{S}(x,Q^2) ~~~~~ \begin{array}{c}
      \\
  \sim \\
  {Q^2 \rightarrow \infty}
\end{array}
~~~~~ C^{(n)}_{S}~\left[{\alpha \over \alpha(t)}\right]^{A^{S}_n},
\ee
where $A^{S}_n = \gamma_n/ 2\pi b$ are the eigenvalues of the matrix (\ref{anomamartix}).
The matrix of anomalous dimensions
for arbitrary n is given by (\ref{quarkgluonanom}),(\ref{quarkgluonanomt1}),
(\ref{quarkgluonanomt3}) and (\ref{quarkgluonanomt2}).

\section{\it Unification of Coupling Constants of Standard Model}
It is interesting to know how the contribution of tensor-gluons changes the high energy
behavior of the coupling constants of the Standard Model \cite{Georgi:1974sy,Georgi:1974yf}.
The coupling constants are evolving in accordance with the formulae
\be\label{system}
{1 \over \alpha_i(M)} = {1 \over \alpha_i(\mu)}+  2 b_i  \ln{M\over \mu},~~~i=1,2,3,
\ee
where we shall consider only the contribution of the lower $s=2$ tensor-bosons:
\be
2b = {58 C_2(G) - 4 n_f T(R) \over 6 \pi}.
\ee
For the $SU(3)_c \times SU(2)_L \times U(1)$  group with its coupling constants $\alpha_3, \alpha_2$ and $\alpha_1$
and six quarks $n_f=6$ and $SU(5)$ unification group we will get
$$
2 b_3= {1 \over 2\pi} 54,~~~ 2 b_2= {1\over 2 \pi} {104\over 3},~~~2 b_1= -{1\over 2 \pi} 4,
$$
so that solving the system of equations (\ref{system}) one can get
\be
\ln{M\over \mu} = {\pi \over 58} \left({1\over \alpha_{el}(\mu)}- {8\over 3} {1\over \alpha_s(\mu)}\right),
\ee
where $\alpha_{el}(\mu)$ and $\alpha_s(\mu)$ are the electromagnetic and strong coupling constants at
scale $\mu$. If one takes $\alpha_{el}(M_Z)= 1/128$ and $\alpha_s(M_Z) =1/10$ one can get that
coupling constants have equal strength at energies of order
$$
M \sim 4 \times 10^4 GeV = 40~ TeV,
$$
which is much smaller than the scale $M\sim  10^{14} GeV$  in the absence of the tensor-gluons contribution.
The value of the weak angle \cite{Georgi:1974sy,Georgi:1974yf} remains intact :
\be
\sin^2\theta_W =  {1\over 6} +{5\over 9} {\alpha_{el}(M_Z)\over \alpha_s(M_Z)},
\ee
as well as the coupling constant at the unification scale remains of the same order
 $\bar{\alpha}(M)=0,01$.

\section{\it Conclusion}

Let us summarize what are the physical consequences of the tensor-gluons hypothesis.
Among all parton distributions, the gluon density $G(x,t)$ is one of the least
constrained functions since it does not couple directly to the
photon in deep-inelastic scattering measurements of the proton $F_2$ structure function.
Therefore it is only indirectly constrained by scaling violations and by the momentum sum rule
which resulted in the fact that only half of the proton momentum is carried by charged
constituents - the quarks -
and that the other part is ascribed to the
neutral constituents. As it was suggested in the
article the process of gluon splitting leads to the emission of tensor-gluons and
therefore a part of the proton momentum which is carried  by the neutral constituents
here is shared between gluons and tensor-gluons.  The density of neutral partons in the proton is
therefore given by the sum of two functions: $G(x,t)+T(x,t)$, where $T(x,t)$ is the density of the
tensor-gluons. To disentangle these contributions and to decide which piece of the neutral partons
is the contribution of gluons and which one
is of the tensor-gluons one should measure the helicities of the neutral
components, which seems to be a difficult task.
The other test of the proposed model will be
the consistency of the mild $Q^2$ behavior of the moments of the structure functions
with the experimental data.

The gluon density can be directly constrained also by jet production \cite{Ellis:1976uc}.
In suggested model the situation is such that the standard quarks cannot radiate tensor-gluons
(such a vertex is absent in the model
\cite{Savvidy:2005fi,Savvidy:2005zm,Savvidy:2005ki,Savvidy:2010vb}),
therefore only gluons are radiated by quarks.
A radiated gluon then can split into a pair of tensor-gluons without obscuring the structure of the
observed three-jet final states.  Thus it seems that there is no obvious contradiction with the existing
experimental data.  Our hypotheses may be wrong, but the uniqueness and simplicity of
suggested extension seems to be the reasons for serious consideration.

In the last section we also observed that the unification scale at which
standard coupling constants are
merging is  shifted to lower energies telling us that it may be that the
new physics is round the corner.
Whether all these phenomena are consistent with experiment is an open question.

This work was supported in part by the
General Secretariat for Research and Technology of Greece and
from the European Regional Development Fund MIS-448332-ORASY (NSRF 2007-13 ACTION, KRIPIS).

\section{\it Note added}
In supersymmetric extensions of the Standard Model \cite{Fayet:1976et,Fayet:1977yc}
the gluons and quarks have natural partners - gluinos of spin s=1/2 and squarks of spin s=0.
If the gluinos appear as elementary constituents of the hadrons then the theory predicts the
existence of new hadronic states, the R-hadrons \cite{Farrar:1978xj,Fayet:1978qc}.
These new hadronic states can be produced in ordinary hadronic collisions and decay
into ordinary hadrons with the radiation of massless photino - the massless partner of
the photon, which takes out a conserved R-parity. Depending on the model  the
gluinos may be massless or
massive, depending on the remaining unbroken symmetries
and the representation content of the theory. The existing experimental data
give evidence that most probably they have to be very heavy
\cite{Farrar:1978rk,Antoniadis:1982qw}.

It seems that phenomenological limitation on the existence of the R-hadrons
is much stronger that the limitation on the existence of the tensor-gluons
inside the ordinary hadrons. This is because gluinos change the statistics of
the ordinary hadrons: the proton has to have a partner - the R-proton which is a boson
and is indeed a new hadron. The existence of tensor-gluon partons inside the proton
does not predict a new hadronic state, a proton remains a proton. The tensor-gluons
change proton dynamical properties - its structure functions.
What are the experimental limitations on that dynamics should
be studied in details.

I would like to thank Pierre Fayet for illuminating discussions of the supersymmetry
phenomenology and physics of R-hadrons and John Iliopoulos
for helpful discussions of the experimental data.

\section{\it Appendix}
The momentum sum rule integrals (\ref{momentumsum}), (\ref{betacoefficient})
and (\ref{tensorbeta}) can be calculated in two different ways: with direct regularization
of the integrals  near $z=1$ and near $z=0$ as it is defined in (\ref{definition}),
or using the substitution $w=1-z$ in the second term of the integrand and then
regularizing  the resulting integrand near $z=1$. As one can convinced both methods
give the identical results. Therefore we shell first calculate the integral over
$P_{TG}(z)$ in (\ref{betacoefficient}) by direct regularization (\ref{definition}) as follows:
\beqa\label{integral1}
& C_2(G)\int_{0}^{1} dz    \left[ {z^{2s+2} \over (1-z)^{2s-1}}
+{(1-z)^{2s+1} \over z^{2s-2} }  \right]   \Rightarrow ~~Reg \nn\\
&= C_2(G)\int_{0}^{1} dz   \left[ {z^{2s+2} \over (1-z)^{2s-1}_+}
+{(1-z)^{2s+1} \over z^{2s-2}_+ }  \right]   =
 - (-1)^{2s}{12 s^2 -1\over 6} C_2(G).
\eeqa\label{integral2}
The same integral we shall calculate now using the substitution $w=1-z$ in the
second term of the integrand and evaluating the sum
\beqa
&  C_2(G)\int_{0}^{1} dz   \left[ {z^{2s+2} \over (1-z)^{2s-1}}
+{z^{2s+1} \over (1-z)^{2s-2} }  \right]
 =C_2(G)\int_{0}^{1} dz    {z^{2s+1} \over (1-z)^{2s-1}}\nn
\eeqa
and then one should regularize the integral near $z=1$ as it is defined in (\ref{definition})
\beqa
& C_2(G)\int_{0}^{1} dz    {z^{2s+1} \over (1-z)^{2s-1}_+}=- (-1)^{2s}{12 s^2 -1\over 6} C_2(G),
\eeqa
which is identical to the previous result.

\end{document}